# A FRAMEWORK FOR PREFETCHING RELEVANT WEB PAGES USING PREDICTIVE PREFETCHING ENGINE (PPE)

Jyoti, A K Sharma, Amit Goel

*Dept. of CE, YMCAUniversity of Sc. & Tech., Haryana, India,*
*justjyoti.verma@gmail.com, ashokkale2@rediffmail.com*

Amit Goel

*Manager, Evalueserve, Gurgaon, Haryana, India,*
*goelamit1@yahoo.com*

Abstract:     This paper presents a framework for increasing the relevancy of the web pages retrieved by the search engine. The approach introduces a Predictive Prefetching Engine (PPE) which makes use of various data mining algorithms on the log maintained by the search engine. The underlying premise of the approach is that in the case of cluster accesses, the next pages requested by users of the Web server are typically based on the current and previous pages requested. Based on same, rules are drawn which then lead the path for prefetching the desired pages. To carry out the desired task of prefetching the more relevant pages, agents have been introduced.

## 1. Introduction

It is indisputable that recent explosion of World Wide Web has transformed not only the discipline of computer-related sciences but also the lifestyles of people and the economies of the countries. Web server is the single most piece of software that enables any kind of web activity. Since its inception, web server has always taken the form of a daemon process. It takes http request, interprets it and serves the file back. As web services are increasingly becoming popular, network congestion and server overloading have become significant problems. To overcome these problems, efforts are being made continuously to increase the web performance.

Web caching is recognized as one of the effective techniques to alleviate the server bottleneck and reduce network traffic, thereby reducing network latency. The basic idea is to cache requested pages at the server so that they don't have to be fetched again. Although web cache schemes reduce the network and I/O bandwidth consumption, they still suffer from a low hit rate, stale data and inefficient resource management. [1] shows that an inefficient web cache management caused a major news web site crash, also called *the Slashdot effect*.

Web prefetch schemes overcome the limitation of web cache mechanisms through pre-processing contents before a user request comes. Web prefetch schemes expect future requests through web log file analysis and prepare the expected requests before receiving it. Compared with web cache schemes, web prefetch schemes focus on the spatial locality of objects when current requests are related with previous requests. Web prefetch schemes increase the bandwidth utilization and reduce or hide the latency due to bottleneck at web server. But prefetching scheme should be carefully chosen as a wrong prefetching system can cause major network bandwidth bottlenecks rather than reducing the web-user-perceived latency.

The organization of the paper is as follows. Section 2 discusses the related work. Section 3 introduces the proposed framework with the required references to the various components of the same. Subsection 3.1 talks about the components of the proposed work followed by the flow process of PPE in subsection 3.2 while subsection 3.3 illustrates the whole process with the help of the flowchart. Section IV concludes the paper followed by the references.

## 2. Related work

Web users can experience response times in the order of several seconds. Such response times are often unacceptable, causing some users to request the delayed documents again. This, in turn, aggravates the situation and further increases the load and the perceived latency. Caching is considered an effective approach for reducing the response time by storing copies of popular Web documents in a local cache, a proxy server cache close to the end user, or even within the Internet. However, the benefit of caching diminishes as Web documents become more dynamic [2]. A cached document may be stale at the time of its request, given that most Web caching systems in use today are *passive* (i.e., documents are fetched or validated only when requested).

Prefetching (or proactive caching) aims at overcoming the limitations of passive caching by proactively fetching documents in anticipation of subsequent demand requests. Several studies have demonstrated the effectiveness of prefetching in addressing the limitations of passive caching (e.g., [3, 4, 5, 6, 7, 8, 9, 10, 11, 12, and 13]). Prefetched documents may include hyperlinked documents that have not been requested yet as well as dynamic objects [14, 11]. Stale cached documents may also be updated through prefetching. In principle, a prefetching scheme requires predicting the documents that are most likely to be accessed in the near future and determining how many documents to prefetch. Most research on Web prefetching focused on the prediction aspect. In many of these studies (e.g., [15, 10]), a *fixed-threshold-based approach* is used, whereby a set of candidate files and their access probabilities are first determined. Among these candidate files, those whose access probabilities exceed a certain prefetching threshold are prefetched. Other prefetching schemes involve prefetching a fixed number of popular documents [9]. Teng et. al [16] proposed the Integration of Web Caching and Prefetching (IWCP) cache replacement policy, which considers both demand requests and prefetched documents for caching based on a normalized profit function. The work in [17] focuses on prefetching pages of query results of search engines. In [18], the authors proposed three prefetching algorithms to be implemented at the proxy server: (1) the *hit-rate-greedy algorithm*, which greedily prefetches files so as to optimize the hit rate; (2) the *bandwidth-greedy algorithm*, which optimizes bandwidth consumption; and (3) the *H/B-greedy algorithm*, which optimizes the ratio between the hit rate and bandwidth consumption. The negative impact of prefetching on the average access time was not considered. Most of the above works rely on prediction algorithms that compute the likelihood of accessing a given file. Such computation can be done by employing Markovian models [19, 10, 20, and 21]. Other works rely on data mining for prediction of popular documents [22] [23] [24] [25]. Numerous tools and products that support Web prefetching have been developed [26], [27, 28, and 29]. Wcol [30] prefetches embedded hyperlinks and images, with a configurable maximum number of prefetched objects. PeakJet2000 [29] is similar to Wcol with the difference that it prefetches objects only if the client has accessed the object before. NetAccelerator [28] works as PeakJet2000, but does not use a separate cache for prefetching as in PeakJet2000. Google's Web accelerator [31] collects user statistics, and based on these statistics it decides on what links to prefetch. It also can take a prefetching action based on the user's mouse movements. Web browsers based on Mozilla Version 1.2 and higher also support link prefetching [32]. These include Firefox [26], FasterFox [33], and Netscape 7.01+ [27]. In these browsers, Web developers need to include html link tags or html meta-tags that give hints on what to prefetch. In terms of protocol support for prefetching, Davison et al. [34] proposed a prefetching scheme that uses a connectionless protocol. They assumed that prefetched data are carried by low-priority datagrams that are treated differently at intermediate routers. Although such prioritization is possible in both IPv6 and IPv4, it is not yet widely deployed. Kokku et al. [35] proposed the use of the TCP-Nice congestion control protocol [36] for low-priority transfers to reduce network interference. They used an end-to-end monitor to measure the server's spare capacity. The reported results show that careful prefetching is beneficial, but the scheme seems to be conservative because it uses an additive increase (increase by 1), multiplicative decrease policy to decide on the amount of data to prefetch. Crovella et. al [37] showed that a rate-control strategy for prefetching can help reduce traffic burstiness and queuing delays. Most previous prefetching designs relied on a *static* approach for determining the documents to prefetch. More specifically, such designs do not consider the state of the network (e.g., traffic load) in deciding how many documents to prefetch. For example, in threshold-based schemes, all documents whose access probabilities are greater than the prefetching threshold are prefetched. As shown in this paper, such a strategy may actually increase the average latency of a document.

## 3. Proposed work

The general architecture of a common Web search engine contains a front-end process and a back-end process. In the front-end process, the user enters the search keywords into the search engine interface, which is usually a Web page with an input box. The application then parses the search request into a form that the search engine can understand, and then the search engine executes the search operation on the index files. After ranking, the search engine interface returns the search results to the user. In the back-end process, a spider or robot fetches the Web pages from the Internet, and then the indexing subsystem parses the Web pages and stores them into the index files. The search engine retrieves the web pages according to the user query. Since *relevancy* is a subjective term, the search results may have varying degree of relevancy for different set of users. Given this fact, there is an opportunity to significantly improve the relevancy of search results for a well defined set of users (example, employees of the same organisation), whose search habbits are largely homogenous.

The proposed work introduces the Predictive Prefetching Engine (PPE) which sits behind the search engine interface. The intent of introducing the PPE [38] is that it will increase the relevancy of the pages returned by the search engine according to the demand of the particular set of users which are termed as *group clients*. PPE also prefetches the pages if it lies in the *rule-database* that is generated by applying the various data mining operations on the *group-client-log*. This log is maintained by the search engine on the request of the various organisations which are assigned a particular set of IP addesses by the Internet Service Providers. The interaction of the PPE with the user and the process of retrieving the relevant web pages from the WWW is explained in the next subsection.

## 3.1. Components of the Proposed Work

1. **Search Engine Interface**: It is the part of the search engine's front end and is basically a web page with the input box. The user enters its query conataining the keywords into this input box and hits the search button.

2. **IP Matcher**: It extracts the IP address from the query coming from a particular user. This IP address is then matched with the particular range of IP addresses for which different *Group-Client-Agents* (GCAs) are defined. Once the GCA is identified, it gets activated.

3. **Group-Client-Agent (GCA)**: As the name suggests, it is an agent. GCA plays the crucial role as it will work on PPE. There are be $n$ GCA's for $n$ group-clients and hence each GCA will have a corresponding PPE to work upon. One group-client refers to a group of users within one organisation. Every organisation is assigned a unique set of IP addresses. These IP addresses will form a part of one group-client.

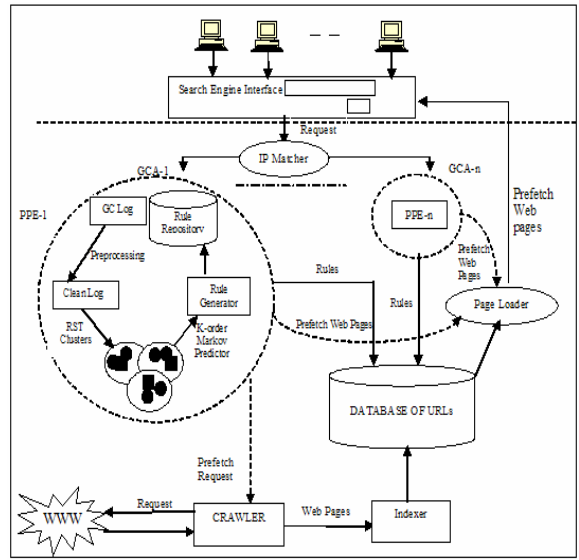

Fig. 1 Framework for retrieving the relevant web pages from WWW using PPE

4. **Group-Client-Log(GC-Log)**: This log is maintained by the search engine on the Group-client's request. The format of the log is same as that of the web server maintained by the search engine and contains every entry from that particular group-client. Each record in the log file contains the client's IP address, the date and time the request is received, the requested object and some additional information such as protocol of request, size of the object etc.

5. **Clean Log:** This log is cleaned by removing all the image files like .jpg and .gif from GC-Log as they yield no productive information about the path followed by the user in a particular session.

6. **RST Clusters**: The clean log is then treated to find the user sessions. A session is the sequence of pages viewed and actions taken by a single user during a defined period of time i.e. 30 minutes. Analyzing the web access log and user sessions, user behavior can be understood. These sessions are then operated upon by the clustering technique known as Rough Set Clustering. The purpose of clustering the sessions is to reduce the search space for applying the various datamining operations. RST operates on the principle of indiscernibility which is defined as equivalence between the objects. RST is chosen as the clustering technique as it aids in decision making in the presence of uncertainty. The result of applying RST is the lower approximation set which contains all the user sessions which definitely contain the target set [39].

7. **Rule Generator**: By making use of rough set clustering, those user sessions were deduced from the web log in which the user spends his quality time. These sessions are in the Lower Approximation set [40]. These sessions are then fed to Rule generator phase of PPE where k-order markov predictors are applied onto these user sessions. It is important to formulate the value of k so that its value is decided dynamically as keeping its value low or high have their own drawbacks. So, the optimum value of k has to be chosen. Here, the minimum threshold that will be used in deciding rules would be half of the maximum number of time a particular sequence of web pages is used. i.e. if the maximum time a particular page sequence called is 6 then minimum threshold to consider other page sequences must be 6/2=3 and k is this minimum threshold. Thus, k is being decided dynamically. The output of this phase is the rules of the form **Di => Dj.**

8. **Rule Repository**: The rules formed in the last phase are then stored in the repository.

9. **Database:** This database contains the URLs of all the pages whose references are stored in the rule repository. The database is enriched by the URLs of rules from all the *n PPE's* by the GCA's..

10. **Page loader:** Its job is to prefetch the pages populated in the hint-list by the GCA onto the client's cache.

### 3.2. FlowProcess for Prefetching the desired documents

Once the IP Matcher identifies the GCA according to the client machine, GCA gets activated and starts working on the prefetching scheme which is as follows:

1. Let the request be for document A.
2. The agent scans the rule database* for the rules of the form A→X for some document X.
3. The agent then scans the database for every rule or part of the rule which has X in its sequence (e.g. A→ Y→X →Z). The only exception to this scan would be in the case of X being the last document in the sequence.
4. As it scans, the agent brings the URLs of all the documents that succeed X from the *Database of*

---

\* The rule database can be organized using some indexing scheme.

*URLs* to its hint list and accordingly prefetch them to the client's cache.

5. The agent continues the scan and populates the hint list till such time the user requests for a web page which doesn't appear in the sequence. In such case, the agent cleans up the hint list and starts afresh. (Step 2).

6. If the GCA finds two rules with the same head but each having a different tail, then the GCA applies the subsequence association rules to find their confidence. The confidence is calculated based on their past history. The rule whose past history generates the maximum confidence is considered by GCA for prefetching. This helps in saving the network bandwidth which is generally considered an issue in the design of the prefetching mechanism.

7. If in case the document A doesn't match as the head of the rule in the *Rule-Repository*, the request is forwarded by the GCA to the crawler. The crawler then crawl the web pages from the WWW and after indexing, add them to the *Database of URLs*.

8. Once the GCA has populated its hint-list with the web pages, it sends the signal to the page loader. The page loader then prefetches the client's cache with the respective GCA's hint list.

### 3.3. Flowchart of the Proposed Work

This subsection discusses the whole process of how the user is returned with the prefetched pages.

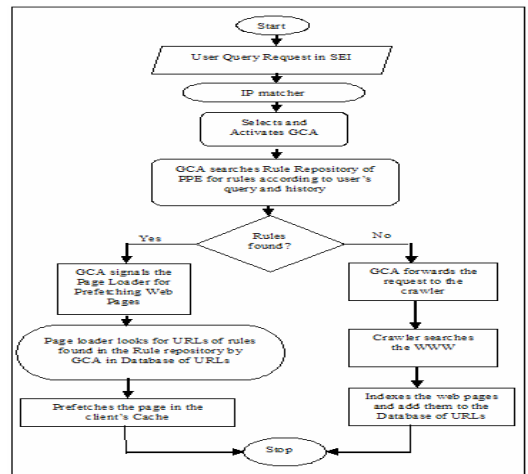

Fig. 2: Flowchart for Prefetching the Web pages according to the rules formed by PPE

Fig. 2 is the abstracted version of the whole process. It is the flowchart that shows the path that is followed right from when the user enters the query in the search engine interface, how a particular GCA (Group-Client-Agent) gets activated and the tasks it then performs to serve its client with the prefetched pages.

# 4. Transaction Processing Phases of PPE

The overall processing of the transactions from the calculation of user accesses to the generation of rules to the Prefetching of the pages into the cache is occurring in three main phases as shown in Fig 3. The step wise working of these phases is as follows:

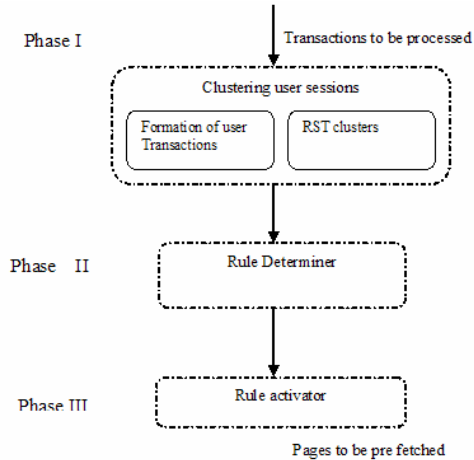

Fig. 3 Transaction Processing Phases

The step wise working of these phases is as follows:

1. Clustering User Sessions: In this phase the user sessions are clustered. To perform this task, two subtasks need to be performed. They are the identification of the user transactions from the GC-Log and then applying RST (Rough Set clustering) over the user sessions to cluster those sessions which definitely contain the target set.

   a. *Identification of the user transactions:* The foremost thing for the determination of the *user transactions* is the identification of the user sessions from the log file. The objective of user session is to separate independent accesses made by different users or by the same user at distant points in time [41, 42].

   b. *RST Clusters:* A rough set, first described by Zdzisław I. Pawlak, is a formal approximation of a crisp set (i.e., conventional set) in terms of a pair of sets which give the *lower* and the *upper* approximation of the original set.

   Formally, an information system is a pair $A = (U, A)$ where U is a non-empty, finite set of objects called the universe and A is a non-empty, finite set of attributes on U. With every attribute $a \in A$, a set $V_a$ is associated such that a: $U \to V_a$. The set $V_a$ is called the domain or value set of attribute a.

   Indiscernibility is core concept of RST and is defined as equivalence between objects. Objects in the information system about which we have the same knowledge form an equivalence relation. The equivalence relation has the following properties.
   If a binary relation $R \subseteq X * X$

   - which is reflexive (i.e. an object is in relation with itself xRx),
   - symmetric (if xRy then yRx),
   - and transitive (if xRy and yRz then xRz)
       is called an equivalence relation.)

   Formally any set $B \subseteq A$, there is associated an equivalence relation called B-Indiscernibility relation defined as follows:
   $IND_A (B) = \{(x, x') \in U^2 \mid \forall a \in B \ a(x) = a(x')\}$

   If $(x, x') \in IND_A (B)$, then objects x and x' are indiscernible from each other by attributes from B.

   Equivalence relations lead to the universe being divided into equivalence class partition and union of these sets make the universal set.

   - Target set is generally supposed by the user.
   - Lower approximation is the union of all the equivalence classes which are contained by the target set. The lower approximation is the complete set of objects that can be *positively* (i.e., unambiguously) classified as belonging to target set *X*.
   - The *P-upper approximation* is the union of all equivalence classes which have non-empty intersection with the target set. It represents the *negative region*, containing the set of objects that can be definitely ruled out as members of the target set.

2. Rule Determiner: Once the user sessions are clustered as lower approximation set, the next step is to determine the rules. These rules will let know which pages are to be prefetched. To determine the rules, markov predictors will be used. E.g. if $S = \langle p_1, \ldots, p_n \rangle$ is a sequence of accesses (called a transaction) made by a user, then the conditional probability that the next access will be $p_{n+1}$ is $P(p_{n+1} \mid p_1, \ldots, p_n )$.

Therefore, given a set of transactions, rules of the form:

$$p_1, \ldots, p_n \Rightarrow p_{n+1} \qquad (1)$$

can be derived, where $P(p_{n+1} \mid p_1, \ldots, p_n)$ is equal to or larger than the user defined cut-off threshold value $T_c$. The left part of the rule is called the *head* and the right part is called the *body*. The body of the rule can also be any length larger than one. E.g. rules of the form

$$p_1, \ldots, p_n \Rightarrow p_{n+1}, \ldots, p_{n+m} \qquad (2)$$

In this case, $P(p_{n+1}, \ldots, p_{n+m} \mid p_1, \ldots, p_n)$, has to be larger than $T_c$.

The dependency of the forthcoming accesses on past accesses defines a *Markov Chain*. The number of past accesses considered in each rule for the calculation of the corresponding conditional probability is called the order of the rule. E.g. the order of the rule *A, B=>C* is 2.

The predictive web prefetching algorithm can be defined as a collection of 1, 2… k-order Markov Predictors. An k-order Markov predictor is defined to be a scheme for the calculation of conditional probabilities $P(p_{n+1}, \ldots, p_{n+m} \mid p_1, \ldots, p_n)$ between document accesses and the determination of the rules of the form (2). The head of the each rule has a size equal to $n$ and the body of each rule has the size equal to $m$.

The job of determining the rules is performed by the *rule generator* component which are then stored in the *Rule repository* component of the proposed framework as shown in Fig.1.

3. Rule Activator: After the determination of the rules of the form (2), the next requirement is for the activation mechanism. The rule activator phase accomplishes the task of finding the prefetched pages from the corresponding rules. This phase makes use of the GCA (group-client agent) which matches the user's request for the documents with the heads of the rules. If the suitable match is found, it will prefetch the documents found in the tail of the corresponding rule.

## 5. Empirical Results

Since relevancy is a very subjective term, to plot it in the form of a graph is difficult. But still keeping some relevancy factor as a baseline, the experimental setup tries to prove the point that

*"For any given set of keywords, no matter how many times they are entered in the search engine within a given time frame, it shows the same position of the URLs that it fetches from the WWW while prefetching if done carefully can greatly help in repositioning the desired URL. "*

For user, to reach from one relevant page to other relevant page, he has to go through all the series of URLs in between. But since the proposed mechanism works over the Rule Repository formed from the past history of the users' access patterns, the rules directly help in prefetching the desired page in the client's cache. Thus, he need not traverse all the in between URLs to reach the desired page.

In the following example, for a particular keyword set "mobile agents based information retrieval for web mining" entered, say the search engine returns 50 URLs in total as noted in table 1. It is known that general search engine returns the most relevant pages on first page with first URL having the highest relevancy and so on. Assuming there are 10 URLs on each page so in total search engine returns 5 pages. The first URL will have highest relevancy with the relevancy factor of 50 and the 50th URL will have the lowest relevancy factor of 1.

**Table 1**

| Keyword | URL No. | Relevancy factor | URL |
|---------|---------|------------------|-----|
| K1 | U1 | 50 | *www.springerlink.com/index/ 3x9vwbd9bxmvcrym.pdf* |
| | U2 | 49 | *portal.acm.org/citation.cfm?i d=1639452* |
| | U3 | 48 | *www.cs.uvm.edu/~xwu/kdd/ WebMining-09.ppt* |
| | U4 | 47 | *www.wikicfp.com/cfp/servlet/ event.showcfp?eventid=1114 4* |
| | U5 | 46 | *scialert.net/fulltext/?doi=jai.2 010.220.238&org=11* |
| | U6 | 45 | *www.sigkdd.org/explorations/ issue2-1/kosala.pdf* |
| | U7 | 44 | …. |
| | … | | … |
| | U50 | 1 | *iaup.org/.../7-summary-report-of-the-international-conference-on-ict-and-knowledge-engineering-2009* |

Table 2 shows the rule set obtained by PPE.

**Table 2**

| S. No | Rules |
|-------|-------|
| 1 | U3->U37 |
| 2 | U2->U12->U18 |
| 3 | U1->U23->U33 |
| 4 | U17->U21 |

Following subsection compares the two approaches:

1. When search engine doesn't perform prefetching.
2. When search engine performs prefetching.

Graph 1

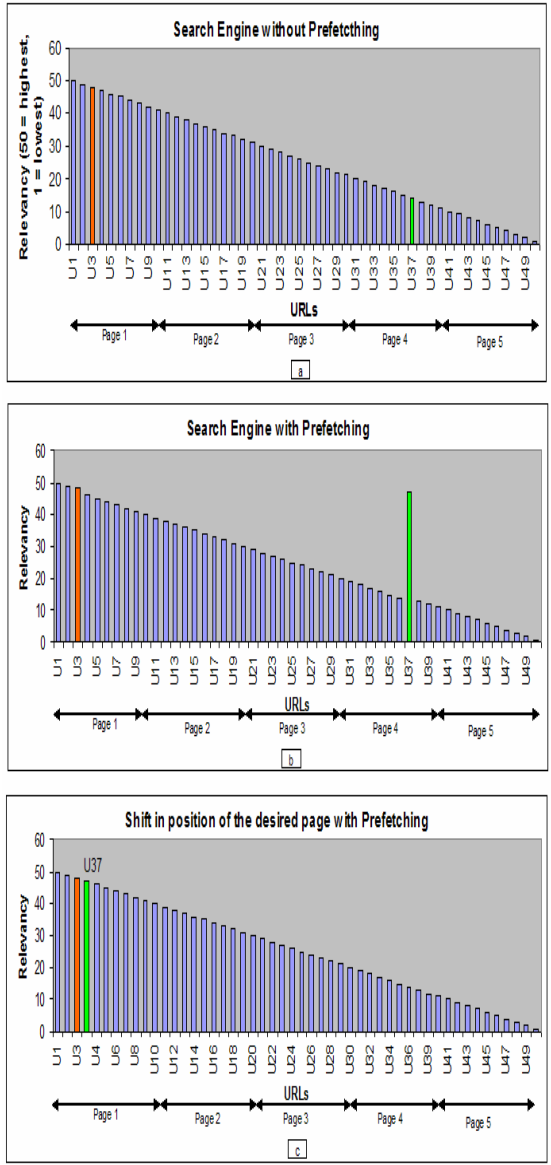

Graph 1 plots the various URLs fetched by the search engine against their relative relevancy when search engines does not employ prefetching. So U1 (URL1) being the most relevant URL according to search engine drags the first position on page 1. Each page contains 10 URLs. So U11 is the first URL on page 2 and U37 is the 7[th] URL on page 3.

Thus, with the increasing page numbers, the relevancy of the URLs decreases according to the normal working of the search engine.

The proposed work on the other hand determines the rules which determine the next likely page to be accessed by the user. It may be observed from the first rule of Table 2, which says that if user accesses U3 then the next URL likely to be accessed by the user is U37. In graph 1(a), the orange bar shows the U3 (the first page accessed by the user according to the keywords entered). For him to access 37, he has to traverse through all the URLs on page 1, page 2 and page 3. Also, according to search engine, the relevancy factor associated with U37 is 14 shown with green bar. Graph 1(b) shows how the relevancy of U37 drastically improves if search engine employs prefetching. The comparison can be seen clearly in graph 1(c) with the relevant orange and green bars coming adjacent to each other.

Thus, U37 which was placed at position 7 on page 3 (position 37 in total listing) by general search engine has moved to position 4 on page 1 when search engine employs prefetching. The following charts 1 & 2 illustrate how the search space for the user reduces drastically if the search engine employs prefetching.

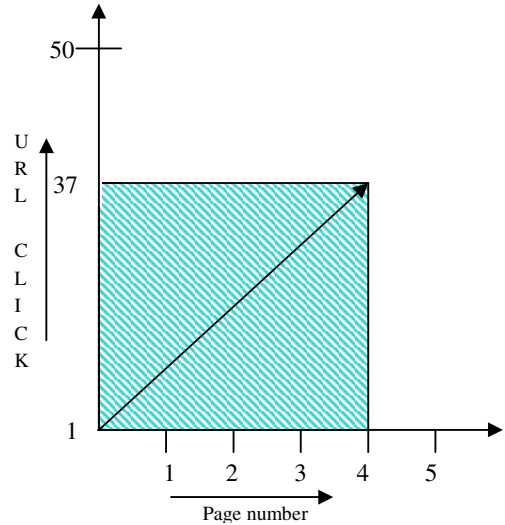

Chart 1: Search space requirement of general search engine

Search area to be covered by user as per results of general search engine is

$$SA= 37*4= 148$$

Search area to be covered by user if search engine employs prefetching is

SA' = 4*1=4

SA'/SA= 4/148= 1/37

If we calculate the ratio of search spaces of both the search engines, it can be seen that search space reduces drastically i.e. by 1/37th, if search engine employs prefetching which proves the point.

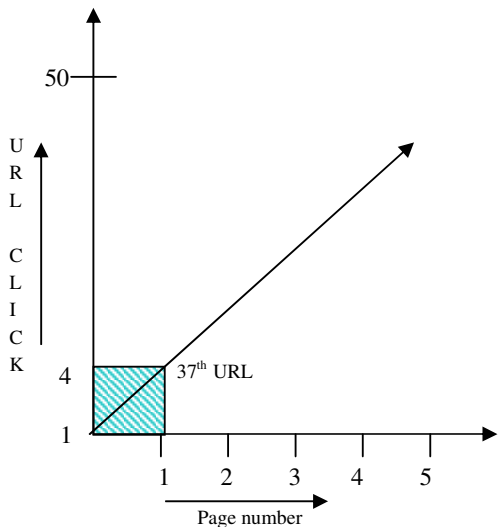

Chart 2: Search space requirement if search engine employs prefetching

CONCLUSION

The search engine retrieves the web pages for the general user. Since *relevancy* is a subjective term, the search results may have varying degree of relevancy for different set of users. The proposed work introduces the PPE for retrieving the web pages for the particular set of users named group-clients whose surfing pattern is logged in the CG-log maintained by the search engine only. Since, these group-clients reflect a particular behaviour over a period of time, PPE encaches the same to return not only the relevant web pages but also prefetches them according to their history. Thus, PPE while prefetching the web pages makes sure that the network bandwidth is not wasted.